\documentclass[aps,prd,twocolumn,showpacs,superscriptaddress,groupedaddress]{revtex4}
\usepackage{dcolumn}   % needed for some tables
\usepackage{bm}        % for math
\usepackage{multirow}

\usepackage{amsmath}
\usepackage{amsfonts}
\usepackage{amssymb}
\usepackage{makeidx}
\usepackage{graphicx}
\usepackage{anysize}

\newcommand{\bs}{\boldsymbol}

\newcommand{\ba}{\begin{eqnarray}}
\newcommand{\ea}{\end{eqnarray}}
\newcommand{\be}{\begin{equation}}
\newcommand{\ee}{\end{equation}}

\usepackage[dvips]{color}

\begin{document}
\title{Global analysis of parity-violating asymmetry data for elastic electron scattering}

\author{R.~Gonz\'alez-Jim\'enez}
\affiliation{Departamento de F\'{\i}sica At\'{o}mica, Molecular y Nuclear, Universidad de Sevilla, 41080 Sevilla, Spain}

\author{J.A.~Caballero}
\affiliation{Departamento de F\'{\i}sica At\'{o}mica, Molecular y Nuclear, Universidad de Sevilla, 41080 Sevilla, Spain}

\author{T.W.~Donnelly}
\affiliation{Center for Theoretical Physics, Laboratory for Nuclear Science and Department of Physics,
Massachusetts Institute of Technology, Cambridge, Massachusetts 02139, USA}

\date{\today}

\begin{abstract}

We perform a statistical analysis of the full set of parity-violating asymmetry data for elastic electron scattering including the most recent high precision measurement from $Q$-weak.
Given the basis of the present analysis, our estimates appear to favor non-zero vector strangeness, specifically, positive (negative) values for the electric (magnetic) strange form factors.
We also provide an accurate estimate of the axial-vector nucleon form factor at zero momentum transfer, $G_A^{ep}(0)$.
Our study shows $G_A^{ep}(0)$ to be importantly reduced with respect to the currently accepted value.
We also find our analysis of data to be compatible with the Standard Model values for the weak charges of the proton and neutron.

\end{abstract}

% PACS numbers:
%
% 12.15.-y  Electroweak interactions
% 12.15.Lk Electroweak radiative corrections %ELIMINADA
% 12.15.Mm Neutral currents %ELIMINADA
% 14.20.Dh Protons and neutrons
% 14.65.Bt Light quarks
% 25.30.Bf Elastic electron scattering
\maketitle

Over the years parity-violating (PV) electron scattering has provided a great deal of precise information on the structure of the nucleon.
A variety of experiments running on different targets, from hydrogen to heavier systems with emphasis on deuterium and helium, have added strong constraints on the electroweak form factors.
Moreover, the high precision reached by the most recent experiments~\cite{PVDIS,Qweak13} will serve as a test of the Standard Model (SM) providing a significant constraint on non-perturbative QCD effects.

The $Q$-weak Collaboration has recently determined the weak charge
of the proton corresponding to the analysis of approximately $4\%$
of the data collected in the experiment~\cite{Qweak13}. The main
objective of the $Q$-weak experiment is to provide a value of
$\sin^2\theta_W$ with a $0.3\%$ precision, that is, the weak charge
of the proton to $4\%$. This extremely small uncertainty will
provide a significant test of the SM. In \cite{Qweak13} a global fit
of data taken for hydrogen, deuterium and helium targets up to
$|Q^2|=0.6$ (GeV/c)$^2$ was also done providing some estimates for
the weak neutral current (WNC) couplings. These results are compatible with the
ones obtained in previous analyses performed by Young and
collaborators~\cite{Young06,Young07}. However, the use of data only
up to $|Q^2|=0.6$ (GeV/c)$^2$ (in the case of \cite{Young06}
only data for $|Q^2|<0.3$ (GeV/c)$^2$ were considered), in addition
to particular assumptions for the $Q^2$-expansion of the form
factors, have convinced us of the necessity of a new and more
complete analysis of the process. Moreover, when such high levels of
precision are the goal, mixing data for elastic scattering on the
proton and $^4$He with those corresponding to the quasielastic (QE)
process, make it difficult to disentangle effects due to the nucleon
structure from others directly linked to final state interactions
(FSI), off-shell effects, few-body nuclear structure, {\it etc.}
Accordingly, in this work we restrict ourselves to elastic electron
scattering processes, and make use of all available data in the
literature with no restriction on the $Q^2$-range considered.

The {\it PV asymmetry} (${\cal A}^{PV}$) in the case of elastic
electron-proton (ep) scattering may be written as
follows~\cite{Gonzalez-Jimenez13a}:
\begin{eqnarray}
{\cal A}^{PV}_{ep} &=& \frac{{\cal A}_0}{2G}\left[
a_A\left(\varepsilon G_E^p\widetilde{G}_E^p+\tau G_M^p\widetilde{G}_M^p\right)\right.\nonumber\\
&-&
\left. a_V\sqrt{1-\varepsilon^2}\sqrt{\tau(1+\tau)}G_M^p G^{ep}_A
 \right] \, ,\label{APVep}
\end{eqnarray}
where $\tau\equiv|Q^2|/(4M^2)$ and $G\equiv \varepsilon(G_E^p)^2+\tau (G_M^p)^2$ with $M$ the nucleon mass and $G_{E,M}^{p}$ the electromagnetic (EM) form factors of the proton.
The term ${\cal A}_0$ determines the scale of the PV asymmetry and is given by ${\cal A}_0=G_F|Q^2|/(2\sqrt{2}\pi\alpha)$ with $G_F$ the Fermi coupling and $\alpha$ the fine structure constant.
Finally, $a_V=-1+4\sin^2\theta_W$ and $a_A=-1$ are the vector and axial-vector WNC electron couplings, and we have introduced the kinematical factor $\varepsilon=[1+2\tau(1+\tau)\tan^2\theta_e/2]^{-1}$ which depends on the scattering angle $\theta_e$.

Assuming charge symmetry, the WNC form factors can be written as follows:
\ba
 \widetilde{G}_{E,M}^p(Q^2) &=& \xi_V^{p}G_{E,M}^{p} +\xi_V^{n}G_{E,M}^{n} +
                                \xi_V^{(0)}G_{E,M}^{s} \nonumber \\
 G^{ep}_A(Q^2)&=&\xi_A^{T=1}G_A^{3} + \xi^{T=0}_A G_A^{8} + \xi_A^{(0)}G_A^{s}\,, \nonumber
 \label{gaXis}
\ea
where $G_{E,M}^s$ are the electric ($E$) and magnetic ($M$) strange form factors and $G_A^{3,8,s}$ are the isovector (3), isoscalar (8) and strange ($s$) contributions to the axial-vector form factor of the proton ($G_A^{ep}$).
The $\xi$ coefficients represent the WNC effective coupling constants that are given in terms
of the weak mixing angle ($\theta_W$) and radiative corrections
(see~\cite{Gonzalez-Jimenez13a,Musolf94} for explicit expressions).

%%%%%%%%%%%
In this work we also consider the data obtained by the HAPPEX
Collaboration on elastic e-$^{4}$He scattering. In this case the PV
asymmetry involves the ratio of the two nuclear monopole form
factors: the EM and the WNC ones. Hence, nuclear effects also play a
role in the description of the process --- for discussions of how
such effects arise, see~\cite{Donnelly89,Moreno09,Moreno13}. It can be shown
that the use of one-body operators (leading-order approximation) and
the assumption that isospin-mixing can be ignored for the $^4$He ground
state yield for the ratio of nuclear form factors simply a ratio of
single-nucleon form factors. 
The latter point was first discussed in 
\cite{Donnelly89} and then revisited in a later study \cite{Ramavataram94} (see, especially, Fig.~2 in that reference). Within the context of the approaches discussed there, the effects of isospin mixing in $^4$He were typically found to be more than an order of magnitude smaller than the effects from strangeness content in the nucleon, in contrast to the expectations for heavier nuclei \cite{Moreno09,Moreno13}. This issue is not completely closed, however, as other approaches find somewhat larger effects from isospin mixing \cite{Viviani07}. In future experimental studies, preferably at low momentum transfers, using a variety of $N=Z$ nuclei it should be possible to address this problem more definitively; for this study we continue to make the above assumptions, 
in which case the PV elastic asymmetry can be written in the form~\cite{Musolf94}
\begin{eqnarray}
{\cal A}^{PV}_{eHe} = -\frac{{\cal A}_0}{2}
  \left[(\xi_V^p+\xi_V^n) + 2\frac{\xi_V^{(0)}G_E^s}{G_E^p+G_E^n}\right]\,.\label{APVhe}
\end{eqnarray}
Thus, these data are also sensitive to the WNC nucleon structure,
and they give information that complements what is obtained through
elastic ep scattering. 

In a recent study~\cite{Gonzalez-Jimenez13a} we showed that at $|Q^2|=1$ (GeV/c)$^2$ (the current limit for asymmetry data) the dispersion in ${\cal A}^{PV}$ due to the use of different prescriptions for the EM form factors (some of them accounting for two-photon exchange contributions) was $\sim3\%$ in the very forward limit ($\theta_e=5^o$) getting much smaller for larger angles and lower $|Q^2|$.
The impact of these uncertainties at the particular Q-weak kinematical conditions has been explored in detail in~\cite{Gonzalez-Jimenez13a}.
Therefore, in the present work we neglect the uncertainties associated with the particular description of the EM nucleon form factors.
On the contrary, ${\cal A}^{PV}$ is highly sensitive to the electric and magnetic strange form factors $G_{E,M}^{s}(Q^2)$ and to the axial-vector one $G_A^{ep}(Q^2)$.
In the latter radiative corrections can introduce very strong effects.
In this work we perform a global fit of the five parameters:
$\xi_V^p$, $\xi_V^n$, $\rho_s$, $\mu_s$ (these two linked to the electric and magnetic strange form factors~\cite{Musolf94}) and the axial-vector form factor at zero momentum transfer: $G_A^{ep}\equiv G_A^{ep}(0)$.
Our predictions, based on a statistical analysis of the full set of PV asymmetry data for elastic ep scattering (SAMPLE~\cite{SAMPLE05}, HAPPEX~\cite{HAPPEX99,HAPPEXa,HAPPEXb,HAPPEXIII}, PVA4~\cite{PVA404,PVA405,PVA409}, G0~\cite{G005,G010} and 
$Q$-weak~\cite{Qweak13}) as well as the two PV elastic e-$^4$He data (HAPPEX~\cite{HAPPEXb,HAPPEXHe}), are compared with all previous analyses presented in the literature.

Assuming the radiative corrections (RC) to be mildly dependent on the energy, namely, on the particular kinematics selected, the 
results of our analysis could serve as a test of their strength. Thus the dispersion found in the values of $\xi_V^p$ and $\xi_V^n$ with 
respect to the corresponding tree-level results (see~\cite{Musolf94} for details) could provide information on $R_V^p$ and $R_V^n$.
Similarly, the deviation in $G_A^{ep}(0)$ would constrain the RC contributions in the axial current as well as possible effects linked to the
anapole moment~\cite{Riska00,Maekawa00a,Maekawa00b}. Recently a large amount of theoretical work has been carried out regarding the 
energy-dependent $\gamma Z$-box correction~\cite{Gorchtein09,Sibirtsev10,Gorchtein11,Hall13}.
Obviously, the results of any global analysis of data are affected by the correction associated with the $\gamma Z$-box. 
However, at present this correction can only be applied to a limited set of data: very forward scattering angles. 
Hence in this work we do not introduce such corrections, so that all data are treated consistently in the global analysis. 
Our present analysis for the weak charges of the proton and neutron, with their uncertainties linked to the parameters that have been estimated, is consistent with the study reported by the Q-weak Collaboration~\cite{Qweak13}, where the $\gamma Z$-box energy-dependent correction was applied.  
However, some caution should be drawn concerning this outcome because of the present limited analysis of Q-weak data (only 4\% of data have been considered). Once the analysis is extended to the 100\% of the Q-weak data, and taking into account that $\xi_V^p$ is basically not constrained by any other data, the $\gamma Z$-box correction to the Q-weak data will become crucial.

Before entering into a detailed discussion of the results, a general comment on the fit procedure should be made.
Specifically, the currently accepted value of the axial-vector form factor at $Q^2=0$, namely, $G_A^{ep}(0) = -1.04\pm0.44$ (see~\cite{Liu07}), has been included as an additional {\it ``experimental''} constraint in the global fit.
Although a significant contribution to $G_A^{ep}$ comes from $G_A^{3}\equiv g_A = 1.2695$ that is well determined from Gamow-Teller $\beta$-decay measurements, radiative corrections can introduce an important uncertainty in $G_A^{ep}(Q^2)$ (see discussion in~\cite{Gonzalez-Jimenez13a}).
Hence there is still room in the global fit procedure for variation of the specific value of $G_A^{ep}(Q^2)$.
In summary, our analysis takes into account all 31 experimental data for the PV asymmetry available in the literature plus the restriction applied to $G_A^{ep}(0)$.

The $Q^2$-dependence in the vector strange and axial-vector form factors has been taken in their usual 
form~\cite{Gonzalez-Jimenez13a,Musolf94}, namely, 
dipole (dipole times $\tau$) for $G_M^s$ ($G_E^s$) with a vector mass $M_V=0.84$ GeV, and dipole shape for the axial form factor
with the axial mass $M_A=1.03$ GeV.
The GKex prescription~\cite{Lomon01,Lomon02,Crawford10} has been used for the EM form factors.
The fit procedure consists in minimizing the $\chi^2$-function:
\ba
\chi^2 = \left[{\bs A_{exp}} - {\bs A_{the}}\right]^T \left[V^{-1}\right]
    \left[{\bs A_{exp}} - {\bs A_{the}}\right]\, ,\label{chi2}
\ea
where ${\bs A_{exp}}$ contains all data and ${\bs A_{the}}$ takes care of the corresponding theoretical predictions that depend on the five parameters considered: $\xi_V^p, \xi_V^n, \rho_s, \mu_s, G_A^{ep}$.
The term $V$ represents the covariance error matrix defined as
\ba
V_{ij} = (\sigma_i^{uncor})^2\delta_{ij} + \sigma_i^{cor}\sigma_j^{cor}\,
\ea
with $\sigma_i^{uncor}$ and $\sigma_i^{cor}$ the uncorrelated and correlated uncertainties of the $i$th-measurement, respectively.
In this work only correlated errors reported by G0 Collaboration are considered.

\begin{table}[htbp]
\centering
\begin{tabular}{ccccc}
\hline
$\xi_V^p$   & $\xi_V^n$ & $G_A^{ep}$ \\
$0.070\pm0.013$  & $-0.946\pm0.044$ &  $-0.62\pm0.41$ \\
\hline \hline
$\rho_s$   & $\mu_s$ \\
$0.92\pm0.58$ & $-0.26\pm0.26$  \\
\hline
\end{tabular}
\caption{Values of the free parameters
from the global fit. The reduced $\chi^2$ value is
$\chi^2_{MIN}/27=1.30$.}
\label{table1}
\end{table}

The results of the $\chi^2$-fit are summarized in Table~\ref{table1} which contains the values for the five free parameters as well as their 1$\sigma$-errors ($\chi<\chi^2_{MIN}+1$).
These values are in good agreement with the SM weak charges of the proton and neutron~\cite{PDG12}: $\xi_V^p=0.0710\pm0.0007$ and $\xi_V^n=-0.9890\pm0.0007$.
On the contrary, $G_A^{ep}$ extracted from the fit,  $G_A^{ep} = -0.62\pm 0.41$, is significantly lower than the value $G_A^{ep} = -1.04\pm0.44$
given in~\cite{Liu07}.
However, note that both results are affected by large errors that make the two predictions overlap. 
These results may reflect important effects coming from higher-order contributions (radiative corrections) as well as from alternative descriptions of the $Q^2$-dependence.
The use of the standard dipole $|Q^2|$-dependence for the axial form factor introduces a systematical uncertainty in the fit that propagates into the
specific value of $G_A^{ep}$ at $|Q^2|=0$. 
% Other functional dependences~\cite{Gonzalez-Jimenez13a} may lead to slightly different results.
Other functional dependences~\cite{Gonzalez-Jimenez13a} as well as the use of a different value of the axial mass $M_A\approx 1.35$ GeV/c as found by the MiniBooNE Collaboration~\cite{MiniBooNENC10} may lead to slightly different results.
Finally, in spite of the significant errors associated with the strange form factors, our global fit favors a large-positive strange charge radius $\rho_s$ (electric strangeness), whereas the strange magnetic moment $\mu_s$ tends to be negative although still being compatible with zero contribution.

\begin{table}[htbp]
\centering
\begin{tabular}{c|ccccc}
\hline
   & $\xi_V^n$ & $G_A^{ep}$ & $\rho_s$  & $\mu_s$ \\
\hline
$\xi_V^p$  & -0.191  & 0.0469 & 0.262 & 0.162 \\
$\xi_V^n$  &  & 0.392 & 0.552 & -0.775 \\
$G_A^{ep}$  &  &  & 0.711 & -0.749 \\
$\rho_s$  &  &  &  & -0.870 \\
\hline
\end{tabular}
\caption{Correlation coefficients between the free parameters of the fit.}
\label{table2}
\end{table}
The correlation coefficients between pairs of the five parameters considered in the $\chi^2$-fit are given in Table~\ref{table2}.
Notice the extremely small correlation for the pair $(\xi_V^p,\, G_A^{ep})$.
This result supports the idea that a determination of the weak charge of the proton is
not affected by the large uncertainty in $G_A^{ep}$.
The correlation values between $\xi_V^p$ and the three remaining parameters, $\xi_V^n$, $\rho_s$ and $\mu_s$, although relatively small, should be taken into account carefully if a high precise determination of the proton weak charge is desired.
To conclude, the rest of parameters: $\xi_V^n$, $G_A^{ep}$, $\rho_s$ and $\mu_s$, are very strongly correlated.

\begin{figure}[htbp]
    \centering
    \includegraphics[width=.45\textwidth,angle=0]{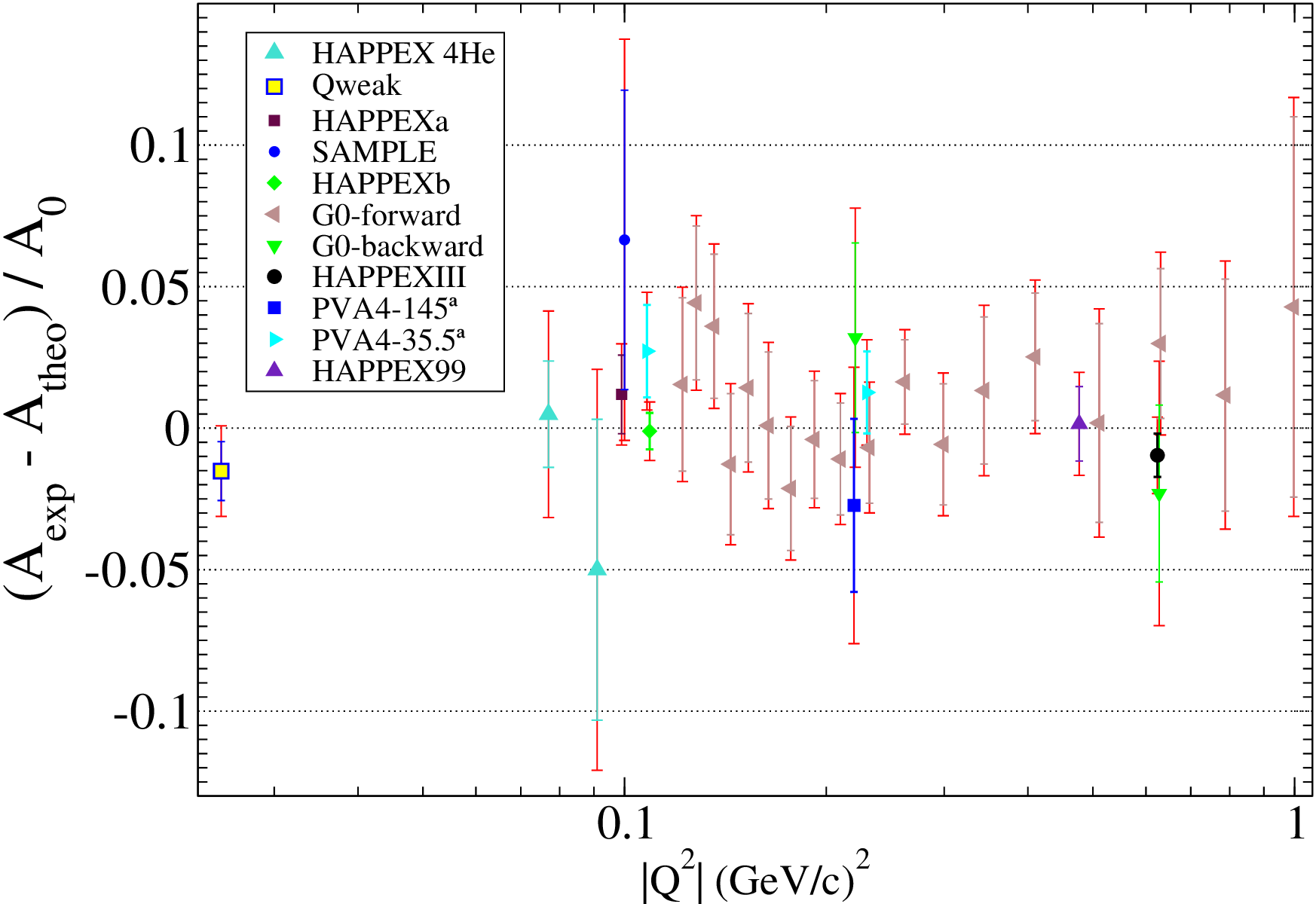}
     \caption{Full set of PV asymmetry experimental data for elastic electron scattering. The data are normalized
       by subtracting our theoretical asymmetry and dividing by ${\cal A}_0$.}
        \label{fig:APVnormalized}
\end{figure}

In Fig.~\ref{fig:APVnormalized} we show all PV asymmetry data for elastic electron scattering compared with our theoretical predictions (zero line).
The inner error bars represent the experimental errors while the outer ones (in red) include the theoretical uncertainty provided by the global fit corresponding to the 1$\sigma$ confidence level.
The strong correlation between some of the parameters has been taken into account in order to determine the theoretical error.
%%%%%%%

\begin{figure}[htbp]
    \centering
    \includegraphics[width=.45\textwidth,angle=0]{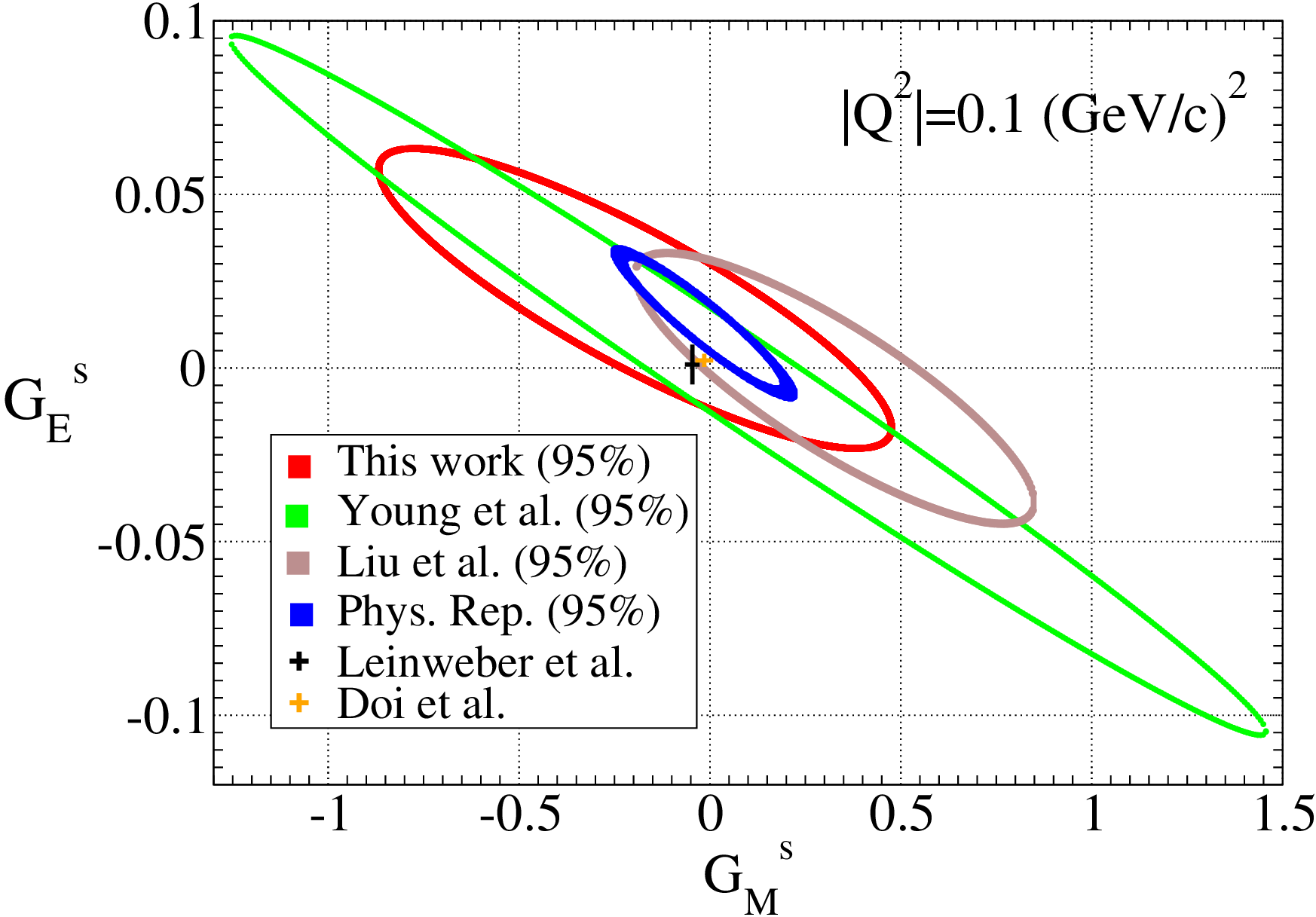}
     \caption{$95\%$ confidence level constraint ellipses
%      ($\chi^2<\chi_{MIN}^2+11.07$) 
     in the plane $G_E^s$-$G_M^s$ at $|Q^2|=0.1$ (GeV/c)$^2$: red $\leftrightarrow$ this work, green $\leftrightarrow$ \cite{Young06}, brown $\leftrightarrow$ \cite{Liu07},
     blue $\leftrightarrow$ \cite{Gonzalez-Jimenez13a}.
     Black and orange crosses are the theoretical predictions presented in~\cite{Leinweber05,Leinweber06} and~\cite{Doi09}, respectively.}
        \label{fig:ges_gms_q2=01}
\end{figure}

In Fig.~\ref{fig:ges_gms_q2=01} we show the 95\% confidence contours obtained in our analysis extrapolated to $|Q^2|=1$ (GeV/c)$^2$.
The results of our present fit (red ellipse) are compared with those from our previous work~\cite{Gonzalez-Jimenez13a} (blue curve).
The significant difference in the areas spanned by the two ellipses is due to the different approaches considered in the two cases.
In~\cite{Gonzalez-Jimenez13a} a global analysis of the full set of PV asymmetry data, including only ep scattering, was performed (28 data) taking only $\rho_s$ and $\mu_s$ as free parameters
(see~\cite{Gonzalez-Jimenez13a} for details).
On the contrary, the red ellipse corresponds to the predictions obtained by simultaneously fitting five parameters.
This larger number of free parameters explains the significant increase in the area.
Note that the present analysis seems to favor positive (negative) values of $\rho_s$ ($\mu_s$).
However, the case of zero strangeness, {\it i.e.,} $\rho_s=\mu_s=0$, is still inside the $95\%$ confidence level in our present analysis (red curve), and only slightly outside the region obtained from our previous study (blue).

We also compare our predictions with the statistical analyses of data performed by
Liu {\it et al.}~\cite{Liu07} (brown ellipse) and Young {\it et al.}~\cite{Young06} (green ellipse).
Also shown for reference the theoretical predictions provided by Leinweber {\it et al.}~\cite{Leinweber05,Leinweber06} (black cross) and Doi {\it et al.}~\cite{Doi09}
(orange cross).
The ellipses corresponding to the works~\cite{Young06,Liu07} are the results of
$\chi^2$-fits of the PV asymmetry data for electron scattering on helium, deuterium and hydrogen in the vicinity of $|Q^2|\approx0.1$ (GeV/c)$^2$.
In particular, in the work of Liu {\it et al.}~\cite{Liu07} a total of 10 data in the range
$0.091\leq|Q^2|\leq0.136$ (GeV/c)$^2$ were employed with only two free parameters: the electric and magnetic strange form factors,
$G_E^s(|Q^2|=0.1$ (GeV/c)$^2$), $G_M^s(|Q^2|=0.1$ (GeV/c)$^2$).
On the contrary, in~\cite{Young06} a total of 19 data in the range $0.038\leq|Q^2|\leq0.299$ (GeV/c)$^2$ were used.
In this case, four free parameters were considered in the fit: the two vector strange
form factors and the axial-vector form factor of the proton and neutron at zero momentum transferred, $G_A^{ep}(0)$ and $G_A^{en}(0)$.

\begin{figure}[htbp]
   \centering
   \includegraphics[width=.45\textwidth,angle=0]{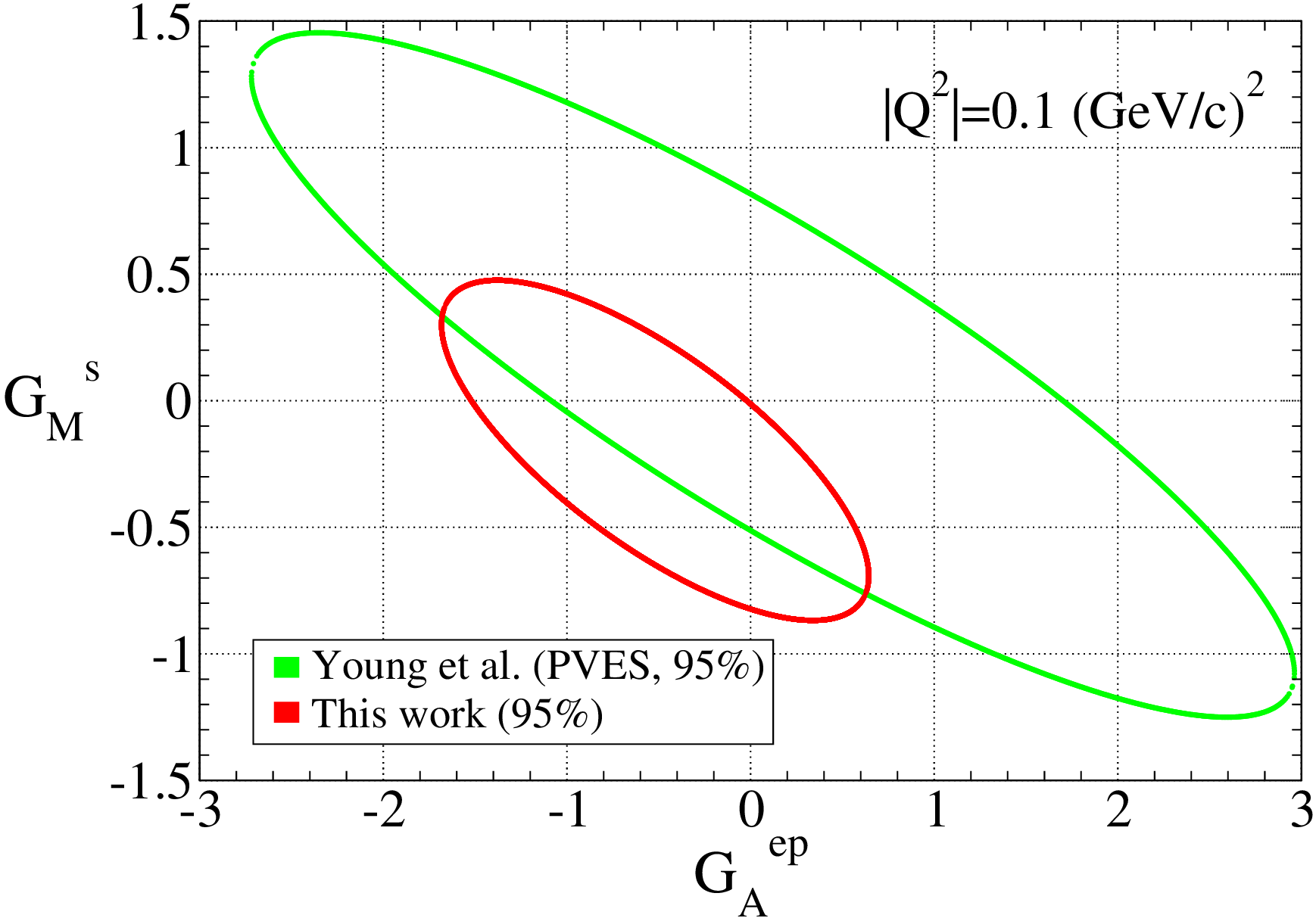}
    \caption{$95\%$ confidence level constraint ellipse in the plane $G_M^s$-$G_A^{ep}$     at $|Q^2|=0.1$ (GeV/c)$^2$: red $\leftrightarrow$ this work, green
        $\leftrightarrow$ \cite{Young06}.}
       \label{fig:gap_GMS_Q2=01}
\end{figure}

In Fig.~\ref{fig:gap_GMS_Q2=01} we present the 95\% confidence level ellipse in the ($\mu_s,G_A^{ep}$) plane.
Results have been extrapolated to the kinematical situation $|Q^2|=0.1$ (GeV/c)$^2$.
Our prediction (red ellipse) is compared with the one of Young {\it et al.}~\cite{Young06} (green ellipse).
As shown, our analysis improves significantly the previous prediction (important reduction in the area of the ellipse).

\begin{figure}[htbp]
    \centering
    \includegraphics[width=.45\textwidth,angle=0]{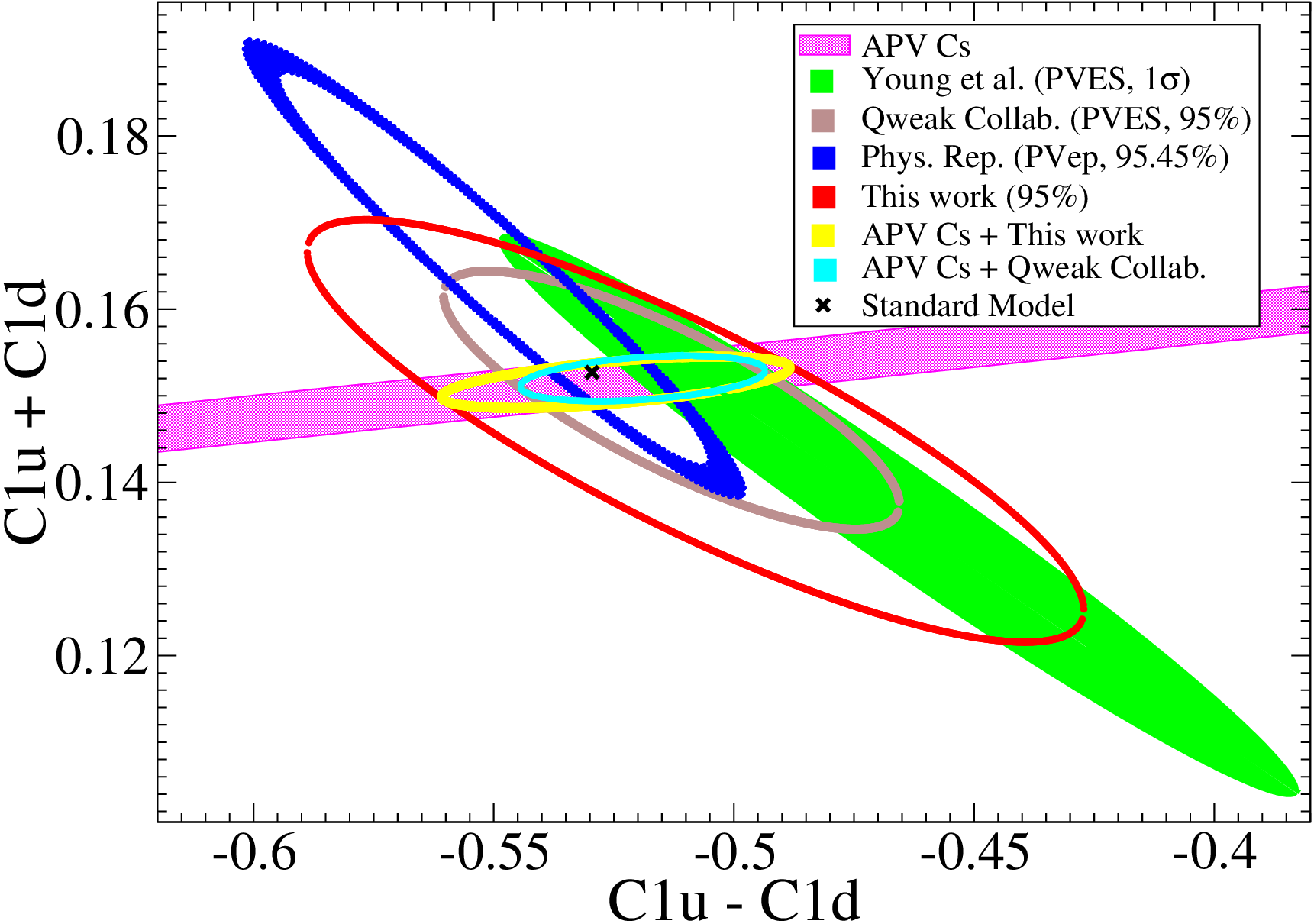}
     \caption{Confidence level ellipses from different works:
     red $\leftrightarrow$ this work,
     blue $\leftrightarrow$ \cite{Gonzalez-Jimenez13a},
     green filled $\leftrightarrow$ \cite{Young07},
     brown $\leftrightarrow$ \cite{Qweak13}.
     The magenta horizontal band represents the constraint from $^{133}Cs$ APV result~\cite{Wood97} (extracted from~\cite{Qweak13}).
%      The yellow ellipse is the result of combining the analysis of this work (red ellipse) and the $^{133}Cs$ APV result.
     The yellow (cyan) ellipse is the result of combining the analysis of this work (Q-weak Collab.~\cite{Qweak13}) and the $^{133}Cs$ APV result.
     The Standard Model prediction is also represented as reference (black cross).}
        \label{fig:C1u_C1d_comp}
\end{figure}

To conclude, we apply our analysis to the determination of the WNC effective couplings.
Thus, we build the 95\% confidence level contour in the plane $(C_{1u}-C_{1d}, C_{1u}+C_{1d})$.
This is shown in Fig.~\ref{fig:C1u_C1d_comp} by the red ellipse.
% The central values and corresponding 1$\sigma$-errors ($\chi<\chi^2_{MIN}+1$) are given by: $C_{1u}-C_{1d} = -0.508\pm0.025$ and $C_{1u}+C_{1d} = 0.146\pm0.008$.
% % with a correlation coefficient $-0.848$.
% The WNC quark coupling constants result:
% $C_{1u}=-0.181\pm0.009$, $C_{1d}=0.327\pm0.015$.
% % with a correlation coefficient $=-0.943$.
Our prediction
is also compared with previous analyses given in the literature:
blue ellipse \cite{Gonzalez-Jimenez13a}, green filled ellipse \cite{Young07}, and brown ellipse \cite{Qweak13}.
The yellow one is the result of combining the constraints coming from our present analysis (red ellipse) with those from Atomic PV in Cesium (APV-Cs) experiments (magenta horizontal band, see~\cite{PDG12,Wood97,Dzuba12} for details).
Likewise, the cyan ellipse is the result of the combined analysis presented in~\cite{Qweak13}.

This combined analysis allows us to get a more accurate determination of the WNC quark coupling constants: $C_{1u}= -0.186\pm0.006$, $C_{1d}= 0.338\pm0.006$
with a correlation coefficient $-0.966$.
Similarly, the weak charges of the proton and neutron result:
$\xi_V^p=0.070\pm0.013$ and $\xi_V^n=-0.978\pm0.012$
with a correlation coefficient $-0.716$.
Our results are in excellent agreeent with the Standard Model predictions~\cite{PDG12}:
$\xi_V^p=0.0710\pm 0.0007$, $\xi_V^n=-0.9890\pm 0.0007$ [$C_{1u}=-0.1885\pm 0.0002$, $C_{1d}=0.3415\pm 0.0002$].

% % % % % % % SUMMARY
Summarizing, we have presented a complete statistical analysis of all PV asymmetry data available for elastic scattering~\footnote{
Although not shown, we have checked with a simple model (static approximation) that the addition to the global analysis of the QE-deuteron
data reported in \cite{SAMPLE04,G010} does not change the conclusions reached in this work. The specific values obtained for the free
parameters are close to the ones shown in Table~I, being within the range shown by the errors.} 
on the proton and on $^4$He.
The $\chi^2$-test is based on the simultaneous fit of five free parameters that are shown to be strongly correlated in most of the cases.
This result may indicate that some caution should be exercised when considering previous analyses that are based on a reduced number
of parameters.
Our new study seems to favor vector strangeness different from zero.
This is consistent with our previous findings~\cite{Gonzalez-Jimenez13a}.
Moreover, a striking result in our analysis is the unexpectedly lower value for $G_A^{ep}$.
However, the axial-vector form factor is known to be highly sensitive to radiative corrections that could explain this result.
In any case, more studies are needed and our prediction could simply be considered as a signal of {\it ``possible''} alternative descriptions for the $Q^2$-dependence of $G_A^{ep}$.
Finally, our results are in accordance with the weak charges of proton and neutron provided by the Standard Model.

\vspace{0.25cm}

This work was partially supported by DGI
(Spain): FIS2011-28738-C02-01, by the Junta de Andaluc\'ia (FQM-160) and
the Spanish Consolider-Ingenio 2000 programmed CPAN,
and in part (TWD) by US Department of Energy under cooperative agreement
DE-FC02-94ER40818.
RGJ acknowledges financial help from VPPI-US (Universidad de Sevilla).

\linespread{0.5}
\appendix
\small
\bibliographystyle{apsrev}

\bibliography{PVES}

\end{document}